\documentclass[article]{revtex4}

\usepackage{amsmath,amssymb,amsthm,amsfonts,mathrsfs,bm}
\usepackage{graphicx,subfigure}

\usepackage[justification=centering]{caption}
\usepackage{amsmath}
\usepackage{graphicx}
\usepackage[colorinlistoftodos]{todonotes}
\usepackage[colorlinks=true, allcolors=blue]{hyperref}
\begin{document}
\title{Superradiant stability of the Kerr black hole}
\author{Wen-Xiang Chen}
\email{wxchen4277@qq.com}
\author{Zi-Yang Huang}
\affiliation{Institute of quantum matter,\\ School of Physics and Telecommunication Engineering,\\
South China Normal University, Guangzhou 510006,China}

\begin{abstract}
We know that Kerr black holes are stable for specific conditions.In this article, we use algebraic methods to prove the stability of the Kerr black hole against certain scalar perturbations. This provides new results for the previously obtained superradiant stability conditions of Kerr black hole. Hod proved that Kerr black holes are stable to massive perturbations in the regime $\mu  \ge \sqrt 2 m{\Omega _H}$. In this article, we consider some other situations of the stability of the black hole in the complementary parameter region$ \sqrt 2  \omega  <  \mu  < \sqrt 2 m{\Omega _H}.$
\end{abstract}
\maketitle

\section{{Introduction}}
Black holes are important and peculiar objects predicted by general relativity. Aspects of black hole physics have been studied extensively. One interesting phenomenon is the superradiance of black holes. Regge and Wheeler \cite{1} proved that the Schwarzschild black holes are stable under disturbance. Due to the phenomenon of superradiance, the stability problem of rotating black holes becomes more complicated. The superradiant effect can occur in classical mechanics and in the scattering prcess of quantum mechanics \cite{2,3,4,5}.

When a  bosonic wave is impinging upon a rotating black hole, the wave reflected by the event horizon will be amplified if the wave frequency $ \omega$ lies in the following superradiant regime\cite{6,7,8,9}
\begin{equation}\label{superRe}
   0<\omega < m\Omega_H ,{{\Omega }_{H}}=\frac{a}{r_{+}^{2}+{{a}^{2}}},
  \end{equation}
where  $m$ is azimuthal number of the bosonic wave mode, $\Omega_H$ is the angular velocity of black hole horizon. This amplification is the superradiant scattering. So through the superradiant process, the rotational energy or electromagnetic energy of a black hole can be extracted. Due to the existence of superradiant modes, a black hole bomb mechanism was proposed by Press and Teukolsky \cite{7}. If there is a mirror between the black hole horizon and space infinity, the amplified wave can be scattered back and forth and grows exponentially, which leads to the superradiant instability of the black hole.

        Although there have been so much study on superradiance of rotating black holes, even Kerr black hole is not investigated thoroughly.Hod proved\cite{10} that the Kerr black hole should be
superradiantly stable under massive scalar perturbation when $\mu \ge \sqrt{2}m\Omega_H$, where $\mu$ is the mass. 

According to\cite{11} , we demonstrate that the Kerr black holes are stable from some other perspectives for certain mass scalar disturbances. We will show that under certain conditions there is only a maximum effective potential outside the event horizon, there is no barrier separated by a potential well, and Kerr black holes are superradiantly stable. In fact, when certain conditions meet, ${\mu} = {y}{\omega}$,
\begin{equation}
\frac{m}{{2\lambda  + \frac{{1 - \lambda }}{2}{y^2} + ({y^2} + 1)\frac{{1 - \lambda }}{{2(1 + \lambda )}}}} < a\omega  < \frac{{\sqrt 2 }}{y}ma{\Omega _H},
\end{equation}
the Kerr black holes are proved to be superradiantly stable in the complementary parameter region
$\mu  < \sqrt 2 m{\Omega _H}$.

\section{{Description of the Kerr-black-hole system}}
The metric of the Kerr black hole\cite{12,13} (in natural unit G=c=1) is
\begin{equation}
d{{s}^{2}}=\frac{\Delta }{{{\rho }^{2}}}{{(dt-a{{\sin }^{2}}\theta d\phi )}^{2}}+\frac{{{\rho }^{2}}}{\Delta }d{{r}^{2}}+{{\rho }^{2}}d{{\theta }^{2}}+\frac{{{\sin }^{2}}\theta }{{{\rho }^{2}}}{{[adt-({{r}^{2}}+{{a}^{2}})d\phi ]}^{2}}.
\end{equation}
\begin{equation}
\Delta =r^{2}-2Mr+a^{2},\rho ^{2}=r^{2}+a^{2}cos^{2}\theta.
\end{equation}
The dynamics of mass scalar field disturbances are controlled by the Klein-Gordon equation
\begin{equation}
({{\nabla }^{\nu }}{{\nabla }_{\nu }}-{{\mu }^{2}})\Psi =0.
\end{equation}

The solution of the above equation with the eigenvalue of the spherical harmonic function \cite{14}can be written as
\begin{equation}
{{\Psi }_{\text{lm}}}\text{(t, r, }\theta \text{, }\phi \text{) }=\text{ }\sum\limits_{\text{l,m}}{{{\text{e}}^{\text{im}\phi }}}{{\text{S}}_{\text{lm}}}\text{(}\theta \text{)}{{\text{R}}_{\text{lm}}}\text{(r)}{{\text{e}}^{\text{-i}\omega \text{t}}}.
\end{equation}
Substituting (6)\ into\ the\ Klein-Gordon\ wave\ equation,\ we\ can\ find\ that\ the\ angular\ function\
${{\text{S}}_{\text{lm}}}\text{(}\theta \text{)}$  satisfies\ the\ angular\ motion\ equation\cite{15,16,17,18,19,20,21,22,23}
\begin{equation}
\frac{1}{\sin \theta }\frac{d(\sin \theta \frac{d{{S}_{lm}}}{d\theta })}{d\theta }+[{{K}_{lm}}+{{a}^{2}}({{\mu }^{2}}-{{\omega }^{2}}){{\sin }^{2}}\theta -\frac{{{m}^{2}}}{{{\sin }^{2}}\theta }]{{S}_{lm}}=0;
\end{equation}
 According to the references\cite{17,18,19,20,21,22}, one key result about the prolate angular eigenvalue is
\begin{equation}
{{K}_{lm}}\ge {{m}^{2}}-{{a}^{2}}({{\mu }^{2}}-{{\omega }^{2}}),
\end{equation}
where\ $l$\ is\ the\ spherical\ harmonic\ index,\ $m$\ is\ the\ azimuthal\ harmonic\ index\ with\
$-l\le m\le l$\ and\ $ \omega$\ is\ the\ energy\ of\ the\ mode.

\section{{The effective binding potential of Kerr-black-hole-massive-scalar-field system}}
The\ radial\ Klein-Gordon\ equation\cite{15,16}\ obeyed\ by\ ${{R}_{lm}}$\ is\ given\ by
\begin{equation}
\Delta \frac{d(\Delta \frac{dR}{dr})}{dr}+UR=0,
\end{equation}
where
\begin{equation}
\Delta ={{r}^{2}}-2Mr+{{a}^{2}},
\end{equation}and
\begin{equation}
U = {[\omega ({{\rm{r}}^2} + {a^2}) - {\rm{m}}a]^2} + \Delta [2{\rm{m}}a\omega  - {\mu ^2}{\rm{(}}{{\rm{r}}^2} + {a^2}) - {K_{lm}}].
\end{equation}

The inner and outer horizons of the black hole are
\begin{equation}
{{r}_{\pm }}=M\pm \sqrt{{{M}^{2}}-{{a}^{2}}},
\end{equation}and it is obvious that
\begin{equation}
{{r}_{+}}+{{r}_{-}}=2M,{{r}_{+}}{{r}_{-}}={{a}^{2}}.
\end{equation}

The superradiant properties of Kerr black holes for mass perturbations can be revealed by the KG equation, and the asymptotic solutions of two extreme radial wave equations are considered under appropriate boundary conditions.We use tortoise coordinate  ${{r}_{*}}$  by equation $ \frac{d{{r}_{*}}^{2}}{d{{r}^{2}}}=\frac{{{r}^{2}}}{\Delta }$\cite{40}\  and\ another\ radial\ function\ $ \psi =\text{rR}$.We get the following radial wave equation

\begin{equation}
\frac{{{\text{d}}^{2}}\psi }{\text{dr}_{*}^{\text{2}}}+V\psi =0,
\end{equation}where
\begin{equation}
V=\frac{U}{{{r}^{4}}}-\frac{2\Delta }{{{r}^{6}}}(Mr-{{a}^{2}}).
\end{equation}

It is easy to obtain the asymptotic behavior of the new potential V as
\begin{eqnarray}
\mathop {\lim }\limits_{r \to {r_ + }} \ V=\frac{[\omega( r_{+}^2+a^2)-ma]^2}{{r_{+}^{4}}},
\mathop {\lim }\limits_{r \to \infty } \ V =\omega^2- {\mu ^2}.
\end{eqnarray}

Then the radial wave equation has the following asymptotic solutions  \cite{24,25,26,27,28,29,30,31,32,33,34,35,36,37,38,39,40}
\begin{equation}
r\to \infty ({{\text{r}}_{*}}\to \infty )\Rightarrow {{R}_{\text{lm}}}\sim \frac{1}{r}{{e}^{-\sqrt{{{\mu }^{2}}-{{\omega }^{2}}}{{r}_{*}}}},
\end{equation}
\begin{equation}
r\to {{r}_{+}}({{r}_{*}}\to -\infty )\Rightarrow {{R}_{\text{lm}}}\sim {{e}^{-i(\omega -m{{\Omega }_{H}}){{r}_{*}}}}.
\end{equation}
When
\begin{equation}
{{\omega ^2} - {\mu ^2}}<0,
\end{equation}
there is a bound state of the scalar field.

When $\varphi=\Delta ^{\frac{1}{2}}R$, radial potential equation(9) can be transformed into the flat space-time wave equation
\begin{equation}
\frac{{{d^2}\varphi }}{{d{r^2}}} + ({\omega ^2} - {V_1})\varphi  = 0,
{V_1} = {\omega ^2} - \frac{{U + {M^2} - {a^2}}}{{{\Delta ^2}}}.
\end{equation}

To see if there are additional potential wells, we analyze the geometry of the effective potential ${{V}_{1}}$. From the following asymptotic behavior of the potential ${{V}_{1}}$:
\begin{equation}
{{V}_{1}}(r\to \infty )\to {{\mu }^{2}} - \frac{{4M{\omega ^2} - 2M{\mu ^2}}}{r}+O(\frac{1}{{{\text{r}}^{2}}}),
\end{equation}
\begin{equation}
{{V}_{1}}(r\to {{\text{r}}_{+}})\to -\infty ,{{V}_{1}}(r\to {{\text{r}}_{-}})\to -\infty ,
\end{equation}
\begin{equation}
{{V}_{1}}^{'}(r\to \infty )\to \frac{{4M{\omega ^2} - 2M{\mu ^2}}}{{{\text{r}}^{2}}}+O(\frac{1}{{{\text{r}}^{3}}}).
\end{equation}

So when
\begin{equation}
{2{\omega ^2} - {\mu ^2}}<0
\end{equation}then
\begin{equation}
{{V}_{1}}^{'}(r\to \infty )<0.
\end{equation}

This means that there may be no potential wells when $ r\to \infty$. In the following, we will show that it has only one extreme value outside the event horizon for ${{V}_{1}}$, no trapping potential exists and Kerr black holes are superradiantly stable.

It was previously proved\cite{10,41}that, for a scalar field of proper mass µ interacting with a spinning Kerr black hole of angular velocity ${{\Omega }_{H}}$, the inequality
\begin{equation}
\mu <\sqrt{2}\text{m}{{\Omega }_{H}}
\end{equation}
provides the upper bound of the fixed Kerr-black hole - massive - scalar field configuration domain.

\section{{The relationship between the roots of the equation (27) and the superradiant stability}}
When $z  =  r - {r_ - }$, the\ explicit\ expression\ of\ the\ derivative\ of\ the\ effective\ potential\cite{11}\ is
\begin{align}
{{V}_{1}}^{'} = \frac{{A{r^4} + B{r^3} + C{r^2} + Dr + E}}{{ - {\Delta ^3}}} = \frac{{{A_1}{z^4} + {B_1}{z^3} + {C_1}{z^2} + {D_1}z + {E_1}}}{{ - {\Delta ^3}}};
\end{align}
\begin{equation}
{A_1} = A;{B_1} = B + 4{r_ - }{A_1};{C_1} = C + (3{r_ - }){B_1} + (6r_ - ^2){A_1};
\end{equation}
\begin{equation}
{D_1} = D + (4r_ - ^3){A_1} + (3r_ - ^2){B_1} + (2{r_ - }){C_1};{E_1} = E + r_ - ^4{A_1} + r_ - ^3{B_1} + r_ - ^2{C_1} + {r_ - }{D_1};
\end{equation}
\begin{equation}
{A_1} = 2M( {\mu ^2}-2{\omega ^2}),
\end{equation}
\begin{equation}
{B_1}=-16Mr_{-}\omega ^2-\mu ^2 r_+^2+2 K _{lm}+3 \mu ^2 r_-^2+2 a^2 \mu ^2,
\end{equation}
\begin{equation}
{C_1}=-24Mr_{-}^{2} \omega ^2 +6\omega a m (r_++r_-)-3(r_+-r_-) \left(a^2 \mu ^2+r_{-}^{2}\mu ^2 +K _{lm} \right),
\end{equation}
\begin{equation}
{D_1}=-16Mr_{-}^{3} \omega ^2 +4aMm (5r_--r_+)\omega+(r_+-r_-)^{2} (r_{-}^{2}\mu ^2 +K _{lm}-1)+a^{2}(\mu ^2(r_+-r_-)^{2}-4m^{2}),
\end{equation}
\begin{equation}
{E_1}=2(a^{2}+r_{-}^{2})^{2}(r_+-r_-) \omega ^2 +4am (a^{2}+r_{-}^{2})(r_--r_+)\omega+2(r_+-r_-)(M^{2}+m^{2}a^{2}-a^{2}).
\end{equation}

From the asymptotic behavior of the two horizons and the infinity effective potential, we know that  $\text{ }{{\text{V}}_{\text{1}}}\text{ }\!\!'\!\!\text{ (z)}=\text{0}$ has at least two roots.As far as the above formula is concerned, it may have four roots with$\text{ }{{\text{V}}_{\text{1}}}\text{ }\!\!'\!\!\text{ (z)}=\text{0}$. We know when certain conditions are met, the equation$\text{ }{{\text{V}}_{\text{1}}}\text{ }\!\!'\!\!\text{ (z)}=\text{0}$ cannot have four positive roots.

For
\begin{equation}
\begin{split}
W={{\text{z}}_{\text{1}}}{{\text{z}}_{\text{2}}}+{{\text{z}}_{\text{1}}}{{\text{z}}_{\text{3}}}+{{\text{z}}_{\text{1}}}{{\text{z}}_{\text{4}}}+{{\text{z}}_{\text{2}}}{{\text{z}}_{\text{3}}}+{{\text{z}}_{\text{2}}}{{\text{z}}_{\text{4}}}+{{\text{z}}_{\text{3}}}{{\text{z}}_{\text{4}}}=\frac{{{C}_{1}}}{{{A}_{1}}},\\
L={{\text{z}}_{\text{1}}}{{\text{z}}_{\text{2}}}{{\text{z}}_{\text{3}}}{{\text{z}}_{\text{4}}}=\frac{{{E}_{1}}}{{{A}_{1}}},
\end{split}
\end{equation}
when\ ${{{A}_{1}}}> 0$,\ if\ ${E_1}>0$,\ ${C_1}<0$,\ then\ $L > 0$\ and\ $W < 0$,\ so\ we\ can\ know\ that\ the\ equation $\text{ }{{\text{V}}_{\text{1}}}\text{ }\!\!'\!\!\text{ (z)}=\text{0}$\ cannot\ have\ more\ than\ two\ positive\ roots(From the fact that L is greater than 0, we know that the positive and negative problems of roots can be divided into three cases: in case 1, all four roots are greater than 0; In case two, only two of the roots are greater than 0; In case three, all four roots are less than 0. And we know from W is less than 0 that only case two is true).

\section{{Find the conditions of the superradiant stability}}
Think\  of\ ${E_1}$ as\  a\  quadratic\   equation, ${E_1} = s{\omega ^2} + t\omega  + p$, so
\begin{equation}
\begin{array}{l}
{E_1} = 4M({a^4} - r_ - ^4){\omega ^2} - 4M{a^3}m\omega  + 12Mam\omega r_ - ^2 - 8{M^2}am\omega {r_ - }\\
 - 4M{a^2} - 4{M^2}{r_ - } + 4{a^2}{r_ - } + 4{M^3} + 4M{a^2}{m^2} - 2M{a^4}{\mu ^2}\\
 + 2M{\mu ^2}r_ - ^4 - 4{a^2}{m^2}{r_ - } - 4{M^2}{\mu ^2}r_ - ^3 + 4{M^2}{a^2}{\mu ^2}{r_ - },
\end{array}
\end{equation}
${\Delta _E} = {t^2} - 4sp,$
if\ ${\Delta _E} < 0$,\ then\ ${E_1} > 0$. We\  find\  it\  that\  ${\Delta _E} < 0$, so ${E_1} > 0$.\ From\ equation (35),\ we\ find\ that\ for\ ${{\text{z}}_{\text{1}}},{{\text{z}}_{\text{2}}}$\ are\ two\ real\ positive\ roots,\ if\ ${{\text{z}}_{\text{3}}},{{\text{z}}_{\text{4}}}$\ are\ two\ real\ roots,\ they\ must\ be\ both\ positive\ or\ both\ negative.

Think\ of\ ${C_1}$\ as\ a\ quadratic\ equation,\ first\ case: when\ ${s_1}<0$,${C_1} = {s_1}{\omega ^2} + {t_1}\omega  + {p_1}$, ${\Delta _C} = {t_1}^2 - 4s{}_1{p_1}$,
if\ ${\Delta _C}< 0$,\ then\ ${C_1} < 0$.

If
\begin{equation}
\frac{M{{m}^{2}}{{r}_{+}}-4M{{\mu }^{2}}r_{-}^{2}({{r}_{+}}-{{r}_{-}})}{2{{r}_{-}}({{r}_{+}}-{{r}_{-}})}<{{K}_{lm}}
\end{equation}
then\ ${{\Delta }_{C}}<0,{C_1}<0$.

For
\begin{equation}
\frac{4M{{\mu }^{2}}r_{-}^{2}({{r}_{+}}-{{r}_{-}})}{2{{r}_{-}}({{r}_{+}}-{{r}_{-}})}\text{=(}{{a}^{2}}+r_{-}^{2}){{\mu }^{2}},
\end{equation}
so inequality (37) can be converted into inequality (39),
\begin{equation}
\frac{M{{m}^{2}}{{r}_{+}}}{2{{r}_{-}}({{r}_{+}}-{{r}_{-}})}<{{K}_{lm}}+\text{(}{{a}^{2}}+r_{-}^{2}){{\mu }^{2}}.
\end{equation}
For $ \lambda =\frac{{{r}_{-}}}{{{r}_{+}}}$, we can obtain the inequality (40) from the inequality (39),
\begin{equation}
\frac{1+\lambda }{4\lambda -4{{\lambda }^{2}}}{{m}^{2}}<{{K}_{lm}}+{{a}^{2}}\text{(1}+\lambda ){{\mu }^{2}}.
\end{equation}
So we can get inequality (41) by dividing the two sides of inequality (40) by ${1+\lambda }$, and
\begin{equation}
\frac{1}{4\lambda -4{{\lambda }^{2}}}{{m}^{2}}-\frac{{{K}_{lm}}}{\text{(1}+\lambda )}<{{a}^{2}}{{\mu }^{2}}
\end{equation}
\text{meets the conditions} of ${{\Delta }_{C}}<0$.
But we see that when ${{K}_{lm}}= {{m}^{2}}-{{a}^{2}}({{\mu }^{2}}-{{\omega }^{2}})$,inequality (41) is not valid in the interval where 
$\mu  < \sqrt 2 m{\Omega _H}$.

Second\ case:\ ${\Delta _C} > 0$, when\ $ \mu  < \sqrt 2 m{\Omega _H}$, if $\omega  < \frac{{{t_1} - \sqrt {{t_1}^2 - 4{s_1}{p_1}} }}{{ - 2{s_1}}}$\, then\ ${C_1} < 0$.

When\ $\frac{1}{{4\lambda  - 4{\lambda ^2}}}{m^2} - \frac{{{K_{lm}}}}{{1 + \lambda }} > {a^2}{\mu ^2} \Rightarrow {\Delta _C} > 0$.

For\cite{10,41,42},\ Kerr\ black\ holes\ are\ stable\ to\ massive\ perturbations\ in\ the\ regime\
$ \mu  \ge \sqrt 2 m{\Omega _H}$,\ and\ when\
 $ \mu  < \sqrt 2 m{\Omega _H}$, ${{{K}}_{lm}}>0$.

We\ know\ that
\begin{equation}
 \frac{{{t_1} - \sqrt {{t_1}^2 - 4{s_1}{p_1}} }}{{ - 2{s_1}}}=\frac{{12Mma - \sqrt {144{M^2}{m^2}{a^2} + 96Mr_ - ^2[12M{\mu ^2}(r_ - ^2 - M{r_ - }) + 6{K_{lm}}({r_ - } - M)]}}}{{48Mr_ - ^2}}.
\end{equation}

For\ ${\mu ^2} = {{y}^2\omega ^2}$, when\ $\mu  < \sqrt 2 m{\Omega _H}$, $\frac{\mu }{{{y}}} < \frac{{{t_1} - \sqrt {t_1^2 - 4{s_1}{p_1}} }}{{ - 2{s_1}}} \Rightarrow {C_1} < 0$, so

 \begin{eqnarray}
12Mma - \frac{{48Mr_ - ^2\mu }}{{{y}}} > \sqrt {144{M^2}{m^2}{a^2} + 96Mr_ - ^2[12M{\mu ^2}(r_ - ^2 - M{r_ - })+ 6{K_{lm}}({r_ - } - M)]}
 \end{eqnarray}
$\Rightarrow {C_1} < 0$.
When\ $ \emph{y} > 2\sqrt 2  \Rightarrow 12Mma - \frac{{48Mr_ - ^2\mu }}{{{y}}} > 0$;

For\ ${K_{lm}} \ge {m^2} - {a^2}({\mu ^2} - {\omega ^2})$, ${m^2} > \frac{{2{\mu ^2}{M^2}r_ + ^2}}{{{a^2}}} \ge 2{a^2}{\mu ^2}$, we\ can\ get\ that\ ${K_{lm}} > (1 + \frac{1}{{y^2}}){a^2}{\mu ^2}$,
so\ the\ (43) formula\ can\ be\ obtained\ from\  the\ following\ formula,

\begin{equation}
\frac{{ - ma}}{{{y}}} + \frac{{2r_ - ^2\mu }}{{y^2}}+\frac{{{a^2}\mu }}{{2M}}(1 + \frac{1}{{y^2}})(M - {r_ - }) > \mu (r_ - ^2 - M{r_ - }).
\end{equation}

Inequality (44) can be converted  into inequality (45) about $ \mu$, and we can make the inequality (45)
\begin{equation}
\mu  > \frac{{ma}}{{\frac{{2r_ - ^2}}{{{y}}} + (M{r_ - } - r_ - ^2){y} + ({y} + \frac{1}{{{y}}})\frac{{{a^2}}}{2M}(M - {r_ - })}}
\end{equation}
set\ up\ (43) formula.

Because the interval we need is $\mu  < \sqrt 2 m{\Omega _H}$, we use the properties of quadratic function to find the range of value about $y$. From inequality (45), we get inequality (46)
\begin{equation}
{y} > \frac{{\sqrt 2 (1 + \lambda ) + \sqrt {2{{(1 + \lambda )}^2} - 4(1 - \lambda )(\lambda  + \frac{{\lambda }}{{1 + \lambda }})(4{\lambda ^2} + \frac{{\lambda (1 - \lambda )}}{{1 + \lambda }})} }}{{2(1 - \lambda )(\lambda  + \frac{{\lambda }}{{1 + \lambda }})}},
\end{equation}
in the end, we know that the\ inequality (45)\ is\ established\  in\ the\ regime\
$\mu  < \sqrt 2 m{\Omega _H}$.

For\ ${\mu ^2} = {{y}^2\omega ^2}$, $\lambda =\frac{{{r}_{-}}}{{{r}_{+}}}$, we\ can\ make\ the\ inequality(45)\ as\ the\ following\ inequality
\begin{equation}
\frac{m}{{2\lambda  + \frac{{1 - \lambda }}{2}{y^2} + ({y^2} + 1)\frac{{1 - \lambda }}{{2(1 + \lambda )}}}} < a\omega.
\end{equation}

Through\ numerical\ analysis,\ when\ ${y} > 4.352 $,\ there\ exists\ a\ certain\ interval\ to\ make\ the\ inequality
\begin{equation}
\frac{m}{{2\lambda  + \frac{{1 - \lambda }}{2}{y^2} + ({y^2} + 1)\frac{{1 - \lambda }}{{2(1 + \lambda )}}}} < a\omega  < \frac{{\sqrt 2 }}{y}ma{\Omega _H}
\end{equation}
set\ up. In the end, inequality (48) can deduce that $ {C_1} < 0$.

For\ ${E_1}>0$, $ {C_1}<0$\ at\ this\ time,\ then\ $ L > 0$ \ and\ $ W < 0$,\ and\ we\ can\ know\ that\ the\ equation\ $ \text{ }{{\text{V}}_{\text{1}}}\text{ }\!\!'\!\!\text{ (z)}=\text{0}$\ cannot\ have\ more\ than\ two\ positive\ roots.\ So\ the Kerr\ black\ hole\ is\ superradiantly\ stable\ at\ that\ time.

\section{{Summary}}
We\ use\ algebraic\ methods\ to\ prove\ that\ stability\ of\ the Kerr\ black\ hole\
against\ certain\ scalar\ perturbations.\ We\ know\ that\ Kerr\ black\ holes\ are\ stable\ to\ massive\ perturbations\ in\ the\ regime\ $ \mu  \ge \sqrt 2 m{\Omega _H}$.\ In\ this\ article, we\ consider\ the\ stability\ of\ Kerr\ black\ hole\ in\ the\ $ \omega$\  interval.

When\
${y} > 4.352$,
\begin{equation}
{(\frac{m}{{2\lambda  + \frac{{1 - \lambda }}{2}{y^2} + ({y^2} + 1)\frac{{1 - \lambda }}{{2(1 + \lambda )}}}})^2} < {a^2}{\omega ^2} < \frac{2}{{{y^2}}}{m^2}\frac{{{\lambda ^2}}}{{{{(1 + \lambda )}^2}}} ,
\end{equation}
we come to the conclusion that the Kerr\ black\ hole\ is\ superradiantly\ stable\ in\ another\ regime.

For example, when\ ${{K}_{lm}}$\ get\ the\ value\ ${{m}^{2}}-{{a}^{2}}({{\mu }^{2}}-{{\omega }^{2}})$, for ${R_1} =  \frac{r}{{{r_ + }}}$,
\begin{equation}
\begin{split}
\begin{array}{l}
{a^2}{V_1} = {a^2}{\omega ^2} - \frac{{{a^2}{\omega ^2}[{R_1}^4(1 - {y^2}) + (\lambda  - {y^2}\lambda ){R_1}^2 + {y^2}{R_1}^3(1 + \lambda ) + {R_1}\lambda (1 + \lambda )] - 2\lambda (1 + \lambda ){R_1}ma\omega  + {m^2}(\lambda {R_1} + {\lambda ^2}{R_1} - \lambda {R_1}^2) + \frac{{(\lambda  - 2{\lambda ^2} + {\lambda ^3})}}{4}}}{{{R_1}^4 - 2{R_1}^3(1 + \lambda ) + {R_1}^2(1 + 4\lambda  + {\lambda ^2}) - 2{R_1}(\lambda  + {\lambda ^2}) + {\lambda ^2}}},
\end{array}
\end{split}
\end{equation}
we can know that the\ trend\ of\ $ {a^2}{V_1}$\ for\ ${R_1}$\ reflect\ the\ trend\ of\ ${V_1}$\ for\  r.

For the interval of inequality(49),\ when\ $y=16$,$ m=1$ and\  $\lambda=0.5$, then\ ${a^2}{\omega ^2}$\ take\ $ 1.2424*{10^{ - 4}}$,$3.3676*{10^{ - 4}}$,$4.4302*{10^{ - 4}}$,\ $5.4928*{10^{ - 4}}$\ and\ $ 7.6180*{10^{ - 4}}$( the\ range\ of\  ${a^2}{\omega ^2}$ \ corresponds\ to\  (49) inequality),
we can get that trend\ of\ ${a^2}{V_1}$ for\ ${R_1}$(look at the FIG. 1).

So we can know that the equation$\text{ }{{\text{V}}_{\text{1}}}\text{ }\!\!'\!\!\text{ (z)}=\text{0}$ cannot have more than two positive roots in the figure, and the results shown in the figure are in good agreement with the conclusions of this article.

 \begin{figure}[htp]
 \centering
\includegraphics[width=15cm,height=7cm]{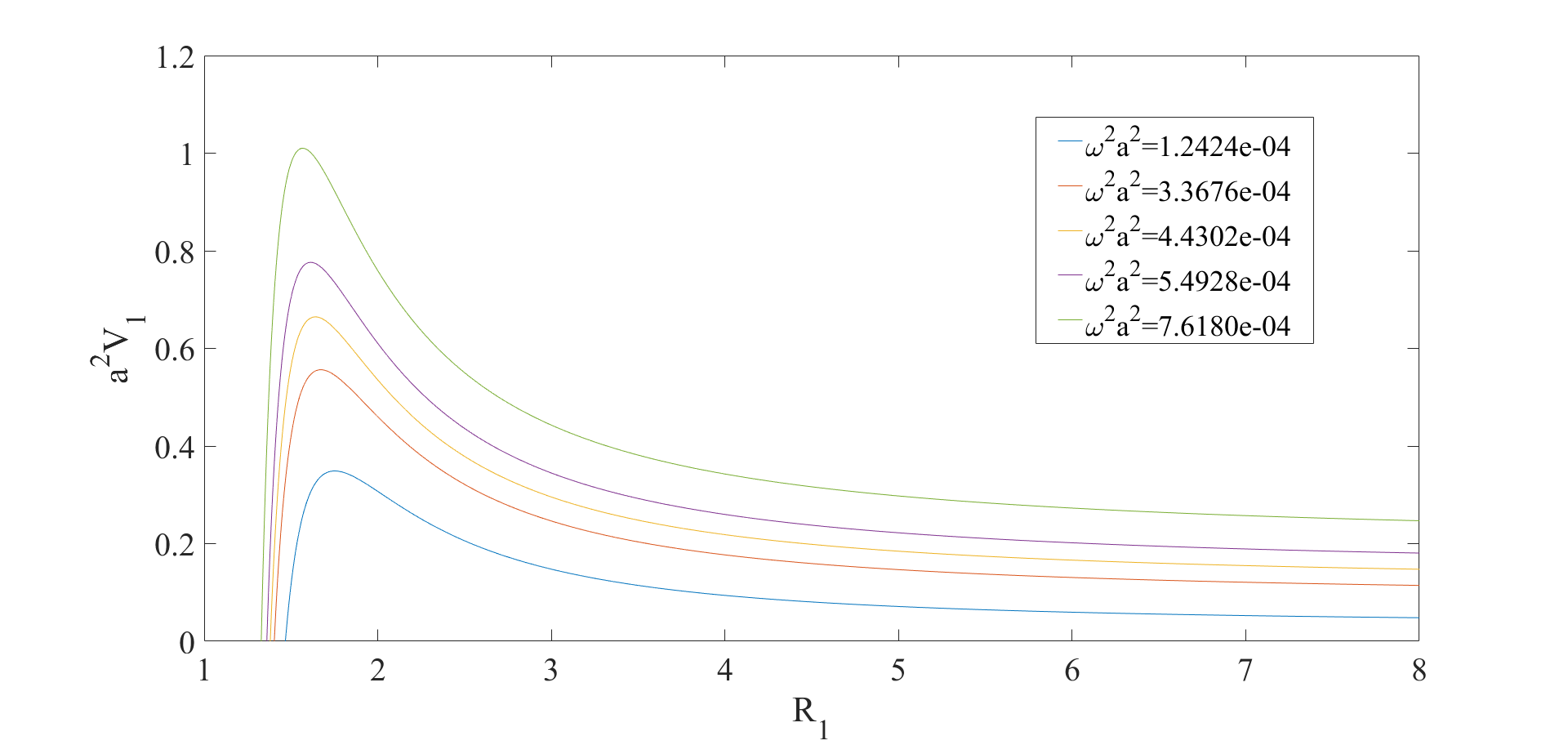}
\caption{ $ {a^2}{V_1}$\ for\ ${R_1}$}
\end{figure}


\begin{thebibliography}{99}
\bibitem[9]{9} We use natural units in which $G = c = \hbar  = 1$.
\bibitem[14]{14} The integer parameters $m$ and  $l \ge \left| m \right|$ are the (azimuthal and spheroidal) harmonic indices of the scalar field modes.
\bibitem[23]{23} It is worth noting that the lower bound on the angular eigenvalues can be saturated in the eikonal $l = m \gg 1$limit, in which case one finds the simple relation .
\bibitem[40]{40}Note that the coordinate $r \in [0,\infty ]$is mapped into $y \in [ - \infty ,\infty ]$ by the radial transformation.
\end{thebibliography}
\end{document}